\documentclass{kapproc} 

\normallatexbib

\begin{document}






\articletitle[]{Time-local master equations: \\ Influence
functional and \\ cumulant expansion}



\author{Heinz-Peter Breuer$^{1,2}$, Andrea Ma$^{2,3}$,
         Francesco Petruccione$^{2,4}$}
\affil{$~^1$ Fachbereich Physik, Carl von Ossietzky Universit\"at, D-26111
             Oldenburg, Germany \\
       $~^2$ Fakult\"at f\"ur Physik, Universit\"at Freiburg,
             D-79104 Freiburg, Germany \\
       $~^3$ Blackett Laboratory, Imperial College, Prince Consort Road, London, England \\
       $~^4$ Istituto Italiano per gli Studi Filosofici, Via Monte di Dio 14,
             I-80132 Napoli}

\email{breuer@marvin.physik.uni-oldenburg.de,
       andrea@www.physik.uni-freiburg.de\\
       petruccione@physik.uni-freiburg.de}







\begin{abstract}
The non-Markovian behaviour of open quantum systems interacting
with a reservoir can often be described in terms of a time-local
master equation involving a time-dependent generator which is not
in Lindblad form. A systematic perturbation expansion of the
generator is obtained either by means of van Kampen's method of
ordered cumulants or else by use of the Feynman-Vernon influence
functional technique. Both expansions are demonstrated to yield
equivalent expressions for the generator in all orders of the
system-reservor coupling. Explicit formulae are derived for the
second and the fourth order generator in terms of the influence
functional.
\end{abstract}




\section{Introduction}
The theory of open quantum systems \cite{TheWork} is concerned
with the elimination of the degrees of freedom of the environment
in order to get an equation of motion for the density matrix of
the reduced system. For an appropriate description of an open
system's dynamics that exhibits non-Markovian features a primary
goal is to derive exact representations for the reduced density
matrix. A closed expression for the density matrix can be obtained
for a few analytically solvable models, such as for the damped
harmonic oscillator and for free Brownian motion
\cite{FeynmanVernon,CALDEIRA,GRABERT}. In many interesting cases,
however, exact representations for the reduced density matrix
serve as starting points of a perturbation expansion in the
system-environment coupling and of the development of numerical
integration schemes.

One possibility of carrying out this program is to derive a
time-local master equation for the open system's density matrix
$\rho_S(t)$ which takes the form
\begin{equation} \label{TCL-MASTER}
 \frac{d}{dt} \rho_S(t) = {\mathcal{K}}(t) \rho_S(t).
\end{equation}
${\mathcal{K}}(t)$ is a time-dependent generator, a super-operator
in the reduced system's Hilbert space ${\mathcal{H}}_S$. As can be
shown with the help of the time-convolutionless (TCL) projection
operator technique \cite{Shibata1,Shibata2} a master equation of
the form (\ref{TCL-MASTER}) indeed exists for small and
intermediate couplings in the case of factorizing initial
conditions. Note that Eq.~(\ref{TCL-MASTER}) is local in time,
i.~e. that it does not involve an integration over the past
history of the reduced system. Due to the explicit time-dependence
of the TCL generator ${\mathcal{K}}(t)$, however, it does not lead
to a quantum dynamical semigroup and, therefore, the generator
need not be in Lindblad form.

A perturbation expansion of the TCL generator ${\mathcal{K}}(t)$
may be found in two different ways. One way is to start from the
formal solution of the von Neumann equation of the total system
and to use van Kampen's technique of the ordered cumulant
expansion \cite{KAMPEN1,KAMPEN2}. Another way is to invoke the
Feynman-Vernon influence functional representation of the reduced
density matrix \cite{FeynmanVernon} and to obtain an expansion of
the generator directly in terms of the influence phase. Both
strategies will be compared in this paper and shown explicitly to
yield identical expansions of the TCL generator.

To be specific we consider a system-reservoir model in which the
reservoir consists of a collection of harmonic oscillators with
frequencies $\omega_n$ and masses $m_n$. The corresponding
coordinates and momenta are denoted by $x_n$ and $p_n$,
respectively. The reservoir Hamiltonian is therefore given by
\begin{equation}
 H_B = \sum_n \left( \frac{1}{2m_n}p_n^2 + \frac{1}{2} m_n \omega_n x_n^2
 \right).
\end{equation}
The system-reservoir coupling is described by the interaction
picture Hamiltonian
\begin{equation} \label{final_ham}
   H_{I}(t) = -\alpha X(t) B(t).
\end{equation}
$\alpha$ represents an overall coupling constant, $X(t)$ is an
interaction picture system operator (not necessarily the position
coordinate) and $B(t)$ the interaction picture reservoir variable
given by
\begin{equation} \label{b(t)}
 B(t)=
 \sum_{n}\kappa _{n}\left( x_{n}\cos \omega _{n}t
 +\frac{p_{n}}{m_{n}\omega _{n}} \sin \omega _{n}t \right).
\end{equation}
The constants $\kappa_n$ describe the strength of the coupling of
the reservoir mode $n$ to the reduced system. The dynamics of the
total system in the interaction picture is then determined by the
von Neumann equation
\begin{equation} \label{eq:Liouville}
 \frac{d}{dt} \rho(t) = \alpha \mathcal{L}(t) \rho(t),
\end{equation}
where we have introduced the Liouville super-operator
$\mathcal{L}(t)$. It is defined by the relation
\begin{equation} \label{defn_liouville}
 \mathcal{L}(t) \rho
 = i \left[ X(t) B(t) ,\rho \right],
\end{equation}
where $\rho$ is any operator of the combined system.

The aim is to eliminate the variables of the reservoir to obtain
an exact representation for the reduced density matrix $\rho_S(t)$
of the open system.  The starting point is the following formal
equation which relates the reduced density matrix $\rho_S(t)$ at
time $t$ to the density matrix $\rho(0)$ of the total system at
the initial time $0$,
\begin{equation} \label{rho_defn}
 \rho_S(t) =
 \mathrm{tr}_B\left\{ {\mathrm{T}}_{\leftarrow}
 \exp \left[ \alpha \int _{0}^{t} dt' \mathcal{L}(t')\right]
 \rho(0) \right\},
\end{equation}
where $\mathrm{tr}_{B}$ stands for the trace over the degrees of
freedom of the environment and $\mathrm{T}_{\leftarrow}$ denotes
the chronological time-ordering operator.

Let us restrict ourselves here to an initial low-entropy state
which is given by a product state of the form
\begin{equation} \label{in_state}
 \rho(0) = \rho_S(0) \otimes \rho_B.
\end{equation}
Here, $\rho_S(0)$ is the density matrix at the initial time and
$\rho_B$ is the density matrix of the reservoir describing a
thermal equilibrium state of temperature $T$ which is given by the
Gibbs state
\begin{equation} \label{therm_equilm_state}
 \rho_B = \frac{1}{Z_{B}} \exp \left(-\beta H_{B}\right),
\end{equation}
where $\beta=1/k_{B}T$ denotes the temperature and $k_{B}$ is the
Boltzmann constant. The normalization factor $Z_B$ represents the
reservoir partition function.

\section{Time-local master equations and ordered cumulants}
We are looking for an appropriate expansion of the TCL generator
of the master equation (\ref{TCL-MASTER}) with respect to the
coupling constant $\alpha$,
\begin{equation} \label{TCL-EXPANSION}
 {\mathcal{K}}(t) = \sum_{n=1}^{\infty} \alpha^n
 {\mathcal{K}}_n(t).
\end{equation}
A general formula for the $n$th-order contribution
${\mathcal{K}}_n(t)$ to the generator of the TCL master equation
can be derived by employing a technique which was developed by van
Kampen for the perturbation expansion of stochastic differential
equations. To explain briefly this method we define for any
super-operator $\mathcal{R}$ of the combined system a
corresponding super-operator $\langle \mathcal{R} \rangle$ of the
reduced system through the relation
\begin{equation}
 \langle \mathcal{R} \rangle \rho_S \equiv \mathrm{tr}_B \left\{
 \mathcal{R} \left( \rho_S \otimes \rho_B \right) \right\}.
\end{equation}
The formal representation (\ref{rho_defn}) may thus be written in
the equivalent form
\begin{eqnarray}
 \rho_{S}(t) &=& \left\langle \mathrm{T}_{\leftarrow }
 \exp \left[ \alpha \int _{0}^{t} dt' \mathcal{L}(t') \right]
 \right\rangle \rho_S(0) \nonumber \\
 &=& \sum_{n=0}^{\infty} \frac{\alpha^n}{n!} \left\langle
 \mathrm{T}_{\leftarrow }
 \left[ \int _{0}^{t} dt' \mathcal{L}(t') \right]^n
 \right\rangle \rho_S(0),
 \label{eq:LiouvilleBrackets}
\end{eqnarray}
where we have expanded the time-ordered exponential in powers of
$\alpha$. Explicitly, the first few terms of this expansion take
the form
\begin{eqnarray} \label{eq:LiouvilleBracketsExplicit}
 \rho_S(t) &=&
 \left[ 1 + \alpha \int_0^t dt_1 \langle {\mathcal{L}}(t_1)
 \rangle + \alpha^2 \int_0^{t} dt_1 \int_0^{t_1} dt_2 \langle
 {\mathcal{L}}(t_1) {\mathcal{L}}(t_2) \rangle \right. \\
 &~& \left. + \alpha^3 \int_0^{t} dt_1 \int_0^{t_1} dt_2  \int_0^{t_2} dt_3
 \langle {\mathcal{L}}(t_1) {\mathcal{L}}(t_2) {\mathcal{L}}(t_3)
 \rangle + \cdots \right] \rho_S(0). \nonumber
\end{eqnarray}
Differentiating this equation with respect to time we get
\begin{eqnarray} \label{eq:LiouvilleBracketsDT}
 \frac{d}{dt} \rho_S(t) &=&
 \left[ \alpha \langle {\mathcal{L}}(t) \rangle
 + \alpha^2 \int_0^{t} dt_1 \langle {\mathcal{L}}(t) {\mathcal{L}}(t_1)
 \rangle \right. \\
 &~& \left. + \alpha^3 \int_0^{t} dt_1  \int_0^{t_1} dt_2
 \langle {\mathcal{L}}(t) {\mathcal{L}}(t_1) {\mathcal{L}}(t_2)
 \rangle + \cdots \right] \rho(0)_S. \nonumber
\end{eqnarray}

The strategy is now to invert the expansion on the right-hand side
of Eq.~(\ref{eq:LiouvilleBracketsExplicit}) with the aim to
express $\rho_S(0)$ in terms of $\rho_S(t)$, and to substitute the
result into Eq.~(\ref{eq:LiouvilleBracketsDT}). As was shown by
van Kampen this procedure can be carried out in a systematic
fashion to yield an expansion for the equation of motion in powers
of the coupling. The result is a time-local master equation of the
form (\ref{TCL-MASTER}), where the $n$th-order contribution to the
TCL generator (\ref{TCL-EXPANSION}) is given by
\begin{equation}
 {\mathcal{K}}_n(t) = \int_0^t dt_1  \int_0^{t_1} dt_2 \ldots
 \int_0^{t_{n-2}} dt_{n-1}
 \langle {\mathcal{L}}(t) {\mathcal{L}}(t_1) {\mathcal{L}}(t_2)
 \ldots {\mathcal{L}}(t_{n-1}) \rangle_{\mathrm{oc}}.
\end{equation}
The quantities
\begin{eqnarray} \label{CUMULANTS-ORDERED}
 && \langle {\mathcal{L}}(t) {\mathcal{L}}(t_1) {\mathcal{L}}(t_2) \ldots
 {\mathcal{L}}(t_{n-1}) \rangle_{\mathrm{oc}} \\
 && \equiv
 \sum (-1)^{(q-1)} \langle {\mathcal{L}}(t) \ldots {\mathcal{L}}(t_i)
 \rangle \langle
 {\mathcal{L}}(t_j) \ldots {\mathcal{L}}(t_k)  \rangle \langle
 {\mathcal{L}}(t_l) \ldots {\mathcal{L}}(t_m) \rangle \langle
 \ldots \rangle \nonumber
\end{eqnarray}
are called ordered cumulants. They are defined by the following
rules. First, one writes down a string of the form
$\langle{\mathcal{L}}\ldots{\mathcal{L}}\rangle$ with $n$ factors
of ${\mathcal{L}}$ in between the brackets. Next one partitions
the string into an arbitrary number of $q$ substrings ($1 \leq q
\leq n$) of the form
$\langle{\mathcal{L}}\ldots{\mathcal{L}}\rangle$ by inserting
angular brackets between the ${\mathcal{L}}$s, whereby each
substring contains at least one factor of ${\mathcal{L}}$. The
resulting expression is multiplied by a factor $(-1)^{(q-1)}$ and
all ${\mathcal{L}}$s are furnished with a time argument in the
following way. The first factor is always ${\mathcal{L}}(t)$. The
remaining ${\mathcal{L}}$s carry any permutation of the time
arguments $t_1,t_2,\ldots,t_{n-1}$ with the only restriction that
the time arguments in each substring must be ordered
chronologically. In Eq.~(\ref{CUMULANTS-ORDERED}) we thus have
\begin{equation}
 t \geq \ldots \geq t_i, \;\;\;
 t_j \geq \ldots \geq t_k, \;\;\;
 t_l \geq \ldots \geq t_m, \;\;\; \ldots
\end{equation}
Finally, the ordered cumulant is obtained by a summation over all
possible partitions into substrings and over all allowed
distributions of the time arguments.

For the thermal state (\ref{therm_equilm_state}) and an
interaction which is linear in the reservoir coordinates and
momenta, as in Eq.~(\ref{final_ham}), we have
\begin{equation}
 \langle \mathcal{L}(t_1) \mathcal{L}(t_2) \ldots
 \mathcal{L}(t_{2n+1}) \rangle = 0.
\end{equation}
Thus, only even-order contributions survive. The above rules then
lead to the following explicit expressions for the second- and the
fourth-order contributions to the TCL generator:
\begin{eqnarray}
  \mathcal{K}_2(t) &=& \int_0^t dt_1 \langle
     \mathcal{L}(t) \mathcal{L}(t_1) \rangle, \label{GEN-K2} \\
  \mathcal{K}_4(t) &=& \int_0^t dt_1  \int_0^{t_1} dt_2
                 \int_0^{t_2} dt_3  \\
        &&   \left( \langle \mathcal{L}(t) \mathcal{L}(t_1)\mathcal{L}(t_2)
               \mathcal{L}(t_3) \rangle
              -\langle \mathcal{L}(t) \mathcal{L}(t_1)
                  \rangle \langle \mathcal{L}(t_2)
               \mathcal{L}(t_3) \rangle \right. \nonumber \\
     & & \left. - \langle \mathcal{L}(t) \mathcal{L}(t_2)
                  \rangle \langle \mathcal{L}(t_1)
               \mathcal{L}(t_3) \rangle
            - \langle \mathcal{L}(t) \mathcal{L}(t_3)
                  \rangle \langle \mathcal{L}(t_1)
               \mathcal{L}(t_2) \rangle \right). \nonumber \label{GEN-K4}
\end{eqnarray}

In general, one expects that a time-local master equation whose
generator consists of only the first few terms of the expansion
provides a good description of the reduced dynamics for weak and
moderate couplings. However, it should be emphasized that an
expansion of the form (\ref{TCL-EXPANSION}) need not exist for
strong couplings. What happens in these cases is that Eq.
(\ref{eq:LiouvilleBrackets}) cannot be solved uniquely for
$\rho_S(0)$. In other words, the initial state $\rho_S(0)$ is not
uniquely determined by the state $\rho_S(t)$ at time $t$. Specific
examples of the application of this technique to physical models
and of the breakdown of the TCL expansion in the strong coupling
regime are discussed in \cite{TheWork}.

\section{Influence functional approach}
In each order of the cumulant expansion the TCL generator involves
certain combinations of $n$-point correlation functions of the
reservoir variables which enter the expressions
(\ref{CUMULANTS-ORDERED}) for the ordered cumulants. Another
strategy of obtaining an expansion of the TCL generator is to
eliminate first the reservoir variables completely form the
expression (\ref{rho_defn}). This is indeed possible for the
present model since the initial state (\ref{in_state}) is Gaussian
with respect to the reservoir variables and since the
system-reservoir interaction is linear in these variables.
Following the procedure used in \cite{QSbremsstrahlung} one finds
the following exact super-operator representation of the reduced
density matrix,
\begin{equation} \label{eq:RhoEx}
 \rho _{S}(t) =
 \mathrm{T}_{\leftarrow }^{X} \exp \left(
 i\alpha^2\Phi_t \left[ X_{c},X_{a}\right] \right)
 \rho _{S}\left( 0 \right),
\end{equation}
where
\begin{eqnarray} \label{INFLUENCE-PHASE}
 i\Phi_t \left[ X _{c},X _{a}\right]
 & = & \int ^{t}_{0} dt' \int^{t'}_{0} dt''
 \left\{ \frac{i}{2}D(t'-t'')
 X _{c}(t') X _{a} (t'')
 \right. \\
 &  & \qquad\qquad\qquad \left. -\frac{1}{2}D_{1}(t'-t'')
 X _{c}(t') X _{c}(t'')
 \right\}. \nonumber
\end{eqnarray}
Equation (\ref{eq:RhoEx}) provides a complete description of the
influence of the reservoir on the reduced system. The motion of
the system is determined by a time-ordered exponential whose
complex phase $i\Phi_t \left[ X _{c},X _{a}\right]$ is a bilinear
functional of the super-operators $X _{c}(t)$ and $X _{a}(t)$. The
action of these super-operators on any matrix $\rho$ is defined
through the commutator and the anti-commutator as
\begin{eqnarray}
X _{c}(t) \rho
   &=&\left[ X (t) ,\rho \right], \\
X _{a}(t) \rho
  &=& \left\{ X (t) ,\rho \right\}.
\end{eqnarray}
$\mathrm{T}_{\leftarrow }^{X}$ denotes the time-ordering of these
super-operators $X_c(t)$ and $X_a(t)$.

The time-ordered exponential function in Eq. (\ref{eq:RhoEx})
represents the super-operator analogue of the Feynman-Vernon
influence functional, which is usually derived utilizing
path-integral techniques \cite{FeynmanVernon,CALDEIRA}. Note that
the double time-integral in Eq. (\ref{INFLUENCE-PHASE}) is already
time-ordered for the integration is extended over the region $t
\geq t' \geq t'' \geq 0$. Two fundamental 2-point correlation
functions enter the above expression for the influence phase,
namely the commutator function
\begin{equation} \label{com_fn}
 D(t-t') \equiv i\left[ B(t),B(t') \right],
\end{equation}
and the anti-commutator function
\begin{equation} \label{anticom_fn}
 D_{1}(t-t') \equiv {\mathrm{tr}}_B \left( \left\{ B(t),
 B(t') \right\} \rho_B \right),
\end{equation}
which are known as dissipation and noise kernel, respectively.

In order to construct a perturbation expansion for a time-local
equation of motion for the reduced density matrix we can proceed
in a similar way as was done for the cumulant expansion: We first
expand the exponential function in Eq. (\ref{eq:RhoEx}),
\begin{eqnarray} \label{expansion}
 \rho_{S}(t) &=&
 \sum ^{\infty }_{m=0}
 \frac{i^m\alpha^{2m}}{m!} \mathrm{T}_{\leftarrow } \left( \Phi_t
 \left[ X _{c},X _{a}\right]
 \right)^{m}\rho _{S}(0) \\
 &=& \left[ 1 + i\alpha^2\Phi_t\left[X _{c},X _{a}\right]
 - \frac{\alpha^4}{2} \mathrm{T}_{\leftarrow }
 \left( \Phi_t\left[X _{c},X _{a}\right]\right)^2 +
 \ldots \right] \rho _{S}(0), \nonumber
\end{eqnarray}
and take the time derivative which leads to the expression:
\begin{eqnarray} \label{t-derivative}
 \frac{d}{dt} \rho_{S}(t) &=& \sum ^{\infty }_{m=1}
 \frac{i^m\alpha^{2m}}{m!} \frac{d}{dt} \mathrm{T}_{\leftarrow } \left( i\Phi_t
 \left[ X _{c},X _{a}\right]
 \right)^{m} \rho _{S}\left( 0 \right) \\
 &=& \left[ i\alpha^2\frac{d}{dt} \Phi_t\left[X _{c},X _{a}\right]
 - \frac{\alpha^4}{2} \frac{d}{dt}  \mathrm{T}_{\leftarrow }
 \left( \Phi_t\left[X _{c},X _{a}\right]\right)^2 +
 \ldots \right] \rho _{S}\left( 0 \right). \nonumber
\end{eqnarray}
A time-local master equation may again be found by solving
Eq.~(\ref{expansion}) for $\rho_S(0)$ within the desired order and
by substituting the result into the right-hand side of
Eq.~(\ref{t-derivative}). It is obvious that this procedure yields
an expansion for a time-local generator of the master equation
which is identical to the one obtained from the cumulant
expansion: The expansions (\ref{eq:LiouvilleBrackets}) and
(\ref{expansion}) are indeed identical to all orders in the
coupling and, therefore, lead to one and the same expansion of the
TCL generator. The comparison of the respective expansions shows
that the relations
\begin{equation} \label{REL-GEN}
 \mathrm{T}_{\leftarrow}^{X} \left( \Phi_t[X_c,X_a] \right)^m
 = \frac{m!}{i^m (2m)!}
 \left\langle \mathrm{T}_{\leftarrow}
 \left[ \int_0^t dt' \mathcal{L}(t') \right]^{2m} \right\rangle
\end{equation}
hold for all $m=1,2,3,\ldots$

Let us illustrate this point by an explicit determination of the
second and the fourth order generator. To lowest order
Eq.~(\ref{expansion}) yields $\rho_S(t)=\rho_S(0)$. To obtain a
second-order equation of motion we thus have to substitute this
lowest-order expression into the right-hand side of
Eq.~(\ref{t-derivative}), keeping only the first term. This leads
to the second-order master equation
\begin{equation}
 \frac{d}{dt}\rho_{S}(t) = {\mathcal{K}}_2(t)
 \rho_S(t)
\end{equation}
with the generator
\begin{equation} \label{eq:InfFunc2Order}
 {\mathcal{K}}_2(t) = i \frac{d}{dt} \Phi_t\left[X _{c},X_{a}\right]
 = \int_0^t dt_1 \langle \mathcal{L}(t) \mathcal{L}(t_1) \rangle,
\end{equation}
where we made use of Eq.~(\ref{REL-GEN}) for $m=1$. Obviously,
this expression coincides with (\ref{GEN-K2}).

The fourth-order contribution of the TCL generator is found by
first inverting Eq.~(\ref{expansion}) in second order which gives
\begin{equation}
 \rho_{S}\left( 0 \right) = \left( 1 - i\Phi_t[X_c,X_a] \right)
 \rho_S(t),
\end{equation}
and by inserting this expression into the right-hand side of
Eq.~(\ref{t-derivative}), keeping only fourth-order terms. The
resulting fourth-order master equation is, obviously,
\begin{equation}
 \frac{d}{dt} \rho_{S}(t)
 = \left[ \mathcal{K}_2(t) + \mathcal{K}_4(t) \right] \rho_S(t),
\end{equation}
where the fourth-order contribution to the generator reads
\begin{equation} \label{K4-PHI}
 \mathcal{K}_4(t) = - \frac{1}{2} \frac{d}{dt}  \mathrm{T}_{\leftarrow }
 \left( \Phi_t\left[X _{c},X _{a}\right]\right)^2
 +\Phi_t\left[X _{c},X _{a}\right]
 \frac{d}{dt}\Phi_t\left[X _{c},X _{a}\right].
\end{equation}
Invoking Eq.~(\ref{REL-GEN}) for $m=1$ and $m=2$ this can be
transformed into
\begin{eqnarray}
 \mathcal{K}_4(t) &=&
 \int_0^t dt_1  \int_0^{t_1} dt_2 \int_0^{t_2} dt_3
 \langle \mathcal{L}(t) \mathcal{L}(t_1)\mathcal{L}(t_2)
 \mathcal{L}(t_3) \rangle \nonumber \\
 &~& -\int_0^t dt_1  \int_0^{t} dt_2 \int_0^{t_2} dt_3
 \langle \mathcal{L}(t) \mathcal{L}(t_1) \rangle
 \langle \mathcal{L}(t_2) \mathcal{L}(t_3) \rangle.
 \label{GEN-K4-INFLUENCE}
\end{eqnarray}
Note that the $t_2$-integral in the second term on the right-hand
side of this equation extends from $0$ to $t$. The triple
time-integral of this term may be brought into time-ordered form
by appropriate substitutions of the time variables. More
precisely, we may write this integral as a sum of three integrals
each of which extends over the region $t \geq t_1 \geq t_2 \geq
t_3 \geq 0$. In this way, one easily recognizes that
Eq.~(\ref{GEN-K4-INFLUENCE}) for the fourth order generator
becomes identical to Eq.~(\ref{GEN-K4}).

By use of Eqs. (\ref{eq:InfFunc2Order}) and (\ref{K4-PHI}) one
easily finds explicit expressions for the second and for the
fourth order contribution to the TCL generator in terms of the
dissipation and the noise kernel and of the commutator and the
anti-commutator super-operators, namely
\[
 {\mathcal{K}}_2(t) =
 \int ^{t}_{0}dt_1 \left\{ \frac{i}{2}D(t-t_1)
 X _{c}(t) X _{a}(t_1)
 -\frac{1}{2}D_{1}(t-t_1) X _{c}(t)
 X _{c}(t_1) \right\},
\]
and
\begin{eqnarray*}
 \mathcal{K}_4(t) &=& \frac{1}{4} \int^{t}_{0}dt_{1}
 \int ^{t_{1}}_{0}dt_{2}\int ^{t_{2}}_{0}dt_{3} \\
 & & \big\{
    \left[ D_{1}(t-t_{2}) D_{1}(t_{1}-t_{3})
    +D_{1}(t-t_{3}) D_{1}(t_{1}-t_{2}) \right] \\
 & & \;\;\;\;\;\times
        X _{c}(t) X _{c}(t_{1})
         X _{c}(t_{2}) X _{c}(t_{3}) \\
 &  & -\left[ D(t-t_{2}) D(t_{1}-t_{3})
      + D(t-t_{3}) D(t_{1}-t_{2}) \right] \\
  & & \;\;\;\;\; \times
      X _{c}(t) X _{c}(t_{1})
      X _{a}(t_{2}) X _{a}(t_{3}) \\
 &  & -i\left[ D_{1}(t-t_{2}) D(t_{1}-t_{3})
      + D(t-t_{3}) D_{1}(t_{1}-t_{2}) \right] \\
   & & \;\;\;\;\; \times
    X _{c}(t) X _{c}(t_{1})
         X _{c}(t_{2}) X _{a}(t_{3}) \\
 &  & -i\left[ D(t-t_{2}) D_{1}(t_{1}-t_{3})
    + D_{1}(t-t_{3}) D(t_{1}-t_{2}) \right] \\
 & & \;\;\;\;\; \times
       X _{c}(t) X _{c}(t_{1})
     X _{a}(t_{2}) X _{c}(t_{3}) \\
 &  & +D(t-t_{2}) D(t_{1}-t_{3})
  X _{c}(t) X _{a}(t_{2})
 X _{c}(t_{1}) X _{a}(t_{3}) \\
 &  & -D_{1}(t-t_{2}) D_{1}(t_{1}-t_{3})
  X _{c}(t) X _{c}(t_{2})
 X _{c}(t_{1}) X _{c}(t_{3}) \\
 &  & +iD(t-t_{2}) D_{1}(t_{1}-t_{3})
  X _{c}(t) X _{a}(t_{2})
 X _{c}(t_{1}) X _{c}(t_{3}) \\
 &  & +iD_{1}(t-t_{2}) D(t_{1}-t_{3})
  X _{c}(t) X _{c}(t_{2})
 X _{c}(t_{1}) X _{a}(t_{3}) \\
 &  & +D(t-t_{3}) D(t_{1}-t_{2})
  X _{c}(t) X _{a}(t_{3})
 X _{c}(t_{1}) X _{a}(t_{2}) \\
 &  & -D_{1}(t-t_{3}) D_{1}(t_{1}-t_{2})
 X _{c}(t) X _{c}(t_{3})
 X _{c}(t_{1}) X _{c}(t_{2}) \\
 &  & +iD(t-t_{3}) D_{1}(t_{1}-t_{2})
 X _{c}(t) X _{a}(t_{3})
 X _{c}(t_{1}) X _{c}(t_{2}) \\
 &  & +iD_{1}(t-t_{3}) D(t_{1}-t_{2})
   X _{c}(t) X _{c}(t_{3})
    X _{c}(t_{1}) X _{a}(t_{2})
 \big\}. \nonumber
\end{eqnarray*}

\section{Conclusion}
The perturbation expansion of the generator of a non-Markovian,
time-local master equation may be constructed through the
technique of ordered cumulants or, equivalently, by use of the
exact influence functional expression for the reduced density
matrix. We have made explicit the relation between both
approaches, invoking the connection between time-ordered products
of the influence phase and of the interaction Liouville operator.

It should be clear, however, that an expansion directly through
the influence functional is only useful, of course, provided an
explicit expression for the influence functional is available.
This was the case in the present study for the reservoir was
assumed to be describable by a Gaussian (thermal) state. In this
respect, the ordered cumulant expansion is more general since it
does not rely on the Gaussian property of the environment.

An important generalization of the present investigation could be
to include non-factorizing initial conditions for the density
matrix of the combined system-reservoir state. The corresponding
correlations in the initial state lead to an inhomogeneity in the
master equation. The expansion of this inhomogeneity in powers of
the system-reservoir coupling is known from the
time-convolutionless projection operator technique
\cite{Shibata1,Shibata2}, while the treatment of non-factorizing
initial conditions is also possible within the framework of the
influence functional technique for Gaussian reservoirs
\cite{GRABERT}.

Finally we emphasize that the derivation of a time-local generator
may be important from the numerical point of view. Not only is a
time-local master equation certainly easier to solve than a
generalized master equation involving a retarded memory kernel,
but it also offers the possibility of a stochastic unraveling of
the master equation: To all orders in the coupling, the form of
the TCL generator allows to design an appropriate stochastic
process $(\phi(t),\psi(t))$ for the state vector in a doubled
Hilbert space ${\mathcal{H}}_S \oplus {\mathcal{H}}_S$ such that
the average over the quantity $|\phi\rangle\langle\psi|$ yields
the open system's density matrix \cite{BKP99}. The TCL form of the
master equation thus gives rise to stochastic wave function
algorithms for non-Markovian quantum processes.










%






\bibliographystyle{apalike}




\end{document}